\documentclass[conference]{worldcomp}

\usepackage[hmargin=.75in,vmargin=1in]{geometry}
\usepackage[american]{babel}
\usepackage[T1]{fontenc}
\usepackage{times}
\usepackage{caption}

\usepackage{textcomp}
\usepackage{epsfig,graphicx}
\usepackage{xcolor}
\usepackage{amsfonts,amsmath,amssymb}
\usepackage{fixltx2e} 
\usepackage{booktabs}

\DeclareMathOperator*{\argmin}{argmin}
\newcommand{\url}[1]{#1}
\newcommand{\K}{\kappa}
\newcommand{\Arabidopsis}{{\em Arabidopsis Thaliana}}

\title{\bf Applications of Machine Learning Methods to Quantifying
Phenotypic Traits that Distinguish the Wild Type from the Mutant
\Arabidopsis\ Seedlings during Root Gravitropism}

\author{
{\bfseries 
Hesam T. Dashti,$^1$ Jernej Tonejc,$^1$ Adel Ardalan,$^2$
Alireza F. Siahpirani,$^3$ Sabrina Guettes,$^1$}\\
{\bfseries Zohreh Sharif,$^4$ Liya Wang,$^5$
Amir H. Assadi$^{1*}$
} \\
$^1$Department of Mathematics, University of Wisconsin, Madison, USA \\
$^2$Database Research Group, Electrical and Computer Engineering
Department, University of Tehran, Iran \\
$^3$Department of Mathematics, University of Tehran, Iran\\
$^4$Department of Mathematics, University of Isfahan, Iran\\
$^5$Cold Spring Harbor Laboratories, NY, USA\\
}

\begin{document}

\maketitle

\footnotetext[1]{
Corresponding author: Amir Assadi (\url{ahassadi@wisc.edu}).\\
All co-authors contributed towards the mathematical computations, design of
algorithms, and development of software. This research is partially
supported by NSF-BIO DBI in 2006-07, and NSF-DMS SCREMS since 2009.
The authors are grateful for helpful scientific discussions in plant
biology and genetics by Professor Patrick Masson. Early parts of this
research has also benefited from discussion of data acquisition and sample
root images from gravitropism experiments by Nathan Miller and Tessa
Durham-Brook images from root branching and growth of lateral roots by
Candace Randall Moore. The authors have benefited from generous
collaboration and technical advice by R\&D staff at nVIDIA Inc., SUN
Network.com (currently SUN-Oracle) and the SUN-Grid computation grant from
Sun Microsystems Inc. The authors gratefully acknowledge the generous help
in engineering of the automation hardware and research collaboration by PBC
Linear Inc.}

\begin{abstract}
Post-genomic research deals with challenging prob\-lems
in screening genomes of organisms for particular
functions or potential for being the targets of genetic engineering for
desirable biological features. 
`Phenotyping' of wild type and mutants is a time-consuming and costly effort by many individuals.
This article is a preliminary progress report in research on large-scale
automation of phenotyping steps (imaging, informatics and data analysis)
needed to study plant gene-proteins
networks that influence growth and development of plants.
Our results undermine the significance of
phenotypic traits that are implicit in patterns of dynamics in plant root
response to sudden changes of its environmental conditions, such as sudden
re-orientation of the root tip against the gravity vector. 
Including dynamic features besides the
common morphological ones has paid off in design of robust and accurate
machine learning methods to automate a typical phenotyping scenario, i.e. to
distinguish the wild type from the mutants.
\end{abstract}

\vspace{1em}


\section{Introduction}
Systems biology has rapidly advanced to offer a number of powerful methods
for discovery of networks of genes/\linebreak[0]proteins and collective
functions of families of genes/\linebreak[0]proteins in life processes.
Analysis of Quantitative
Traits Loci (QTL) refers to a systematic method for discovery of genes and
their functions in systems biology. The major bottleneck in QTL progress
appears to be in the more general and challenging question of how to
quantify phenotypic traits, such as morphology and dynamics of its
variation. QTL and other gene function discovery methods rely on
quantification of phenotypic traits that could be observed and used in a
systematic way to distinguish between organisms with differing DNA
sequences or epigenetic signatures. To extract phenotypic traits that
distinguish genotypic characteristics, biologists make repeated
observations of the wild type and the mutants of the same species during
growth, behavior or in the course of response to external stimuli. In the
following, we illustrate and outline the steps for identifying and
quantifying phenotypic traits of seedling of the model plant
\Arabidopsis\ subject to changes in gravitational force relative to its
natural root orientation during a normal course of growth. It will be
demonstrated that the selected quantitative traits are together
sufficiently informative to carry genomic signatures that are at work
differently in the wild type and mutant plants.

\section{Background and Biological Significance}
Plant roots vary a great deal in morphology, size and complexity of their
architecture. The model plant \Arabidopsis\ has a root system that
is exemplary in both prevalence of its morphological pattern and its
relative simplicity. There are several visible morphological features of
this root system. The primary root is the main branch that starts its life
immediately after germination. Besides being a significant method for
simplifying morphological measurement, midline extraction serves as an
example of high-dimensional data transformation that achieves
dimensionality reduction with some loss of information beyond noise,
namely, ignoring the signal-parameters underlying phenotypic variation such
as morphology of the root due to image texture or image structures such as
shadows of root hair. In higher resolution images, the interpretation will
change, and root hairs could also be part of the features whose
morphological diversity and variation in growth dynamics become essential
biological quantities to be studied that are related to specific gene
functions, pathways or gene-protein network dynamics. In this article we
provide details of a set of algorithms for extraction of root and hair
growth information.

Tropism refers to the directed growth responses of plants to external
stimuli such as gravity, water, light, and temperature. Beginning with the
pioneering work of Darwin, the study of tropism has grown to a major area
of research in plant biology in the course of its century-long history.
Darwin observed that grass seedlings grown in dark tend toward a light
source when illuminated from one side. Plant roots show a similar response
to gravity. When a seed germinates, the roots penetrate the soil and grow
downward. However, if a root is reoriented by 90$^\circ$ with respect to the
gravitational field, the root responds by altering its direction of growth,
curving until it is again vertical. Sachs was the first one to propose a
quantitative measure for gravitropism \cite{JS1887}, namely, the
gravitropic response was proportional to the component of the gravity
vector perpendicular to the root axis. Early studies, using maize roots,
demonstrated distinct regions along the root axis with different
physiological response patterns. Despite this long and illustrious research
history, the molecular mechanisms involved in sensing the gravitational
signal and its transduction still need to be studied. Other tropic
responses are sources of gaining information about the physiological and
molecular processes that influence the tropic response, e.g. the plant
hormone auxin has been implicated in tropic responses through extensive
research by a broad group of scientists (\cite{sg2008,bss2009,mb2001}).

Auxin is involved in asymmetric tropic growth, vascular development and root
formation, as well as in a plethora of other processes. The approaches by
which auxin has been implicated in tropisms include isolation of mutants
altered in auxin transport or response with altered gravitropic or
phototropic response, identification of auxin gradients with radiolabeled
auxin and auxin-inducible gene reporter systems, and by use of inhibitors
of auxin transport that block gravitropism and phototropism. Proteins that
transport auxin have been identified and the mechanisms which determine
auxin transport polarity have been explored. The mechanisms of auxin
action in the gravitropic response and phototropism have recently been
revealed by the analysis of mutants that are defective in the response to
the gravity signal, or have a characteristic response to blue and other
spectral bands of light. Gravitational stimulus induces curvature of the
primary root of \Arabidopsis. It is known that the patterns of
root curvature reveal quantitative traits that are associated to a number of
genes and proteins that play vital roles in growth and development of the
plant.

\section{Instrumentation and Methods}

The application described below gives an outline of machine-learning
methods that distinguish mutant from wild type seedling plants in gravitropism
experiments. We note that these applications could be adapted
for other experimental protocols in plant biology that attempt to quantify
subtle phenotypic traits in order to decipher functions of genes-proteins.
 
\subsection{Automation and High Throughput Imaging} 
\begin{figure}[htb]\centering
\includegraphics[width=\columnwidth]{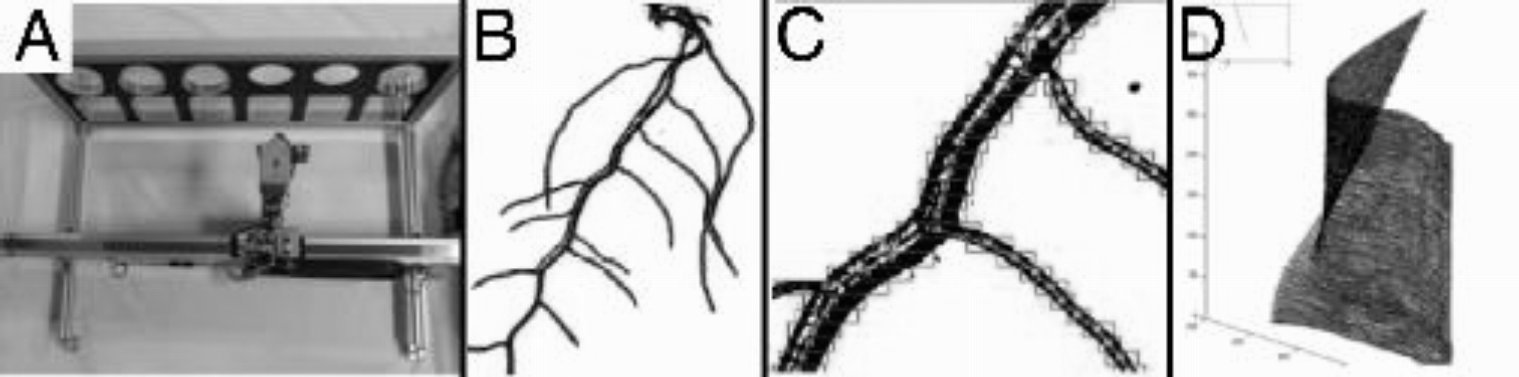}
\caption{\small [A] Top view of the experimental fully automated imaging device
engineered in collaboration with PBC Linear Co. [B,C] Parts of a high
resolution image of growth of lateral roots and extraction of morphological
through the algorithms developed by the authors. [D] The midlines of primary
root images frames (please see below) could be metaphorically positioned on
top of each other to visualize the dynamics of growth in gravitropism as
the geometry of the surface. We have developed algorithms for extraction of
midlines of root hair and dynamics of growth as the next step in
capturing phenotypic variation.} \label{fig1}
\end{figure}

We have designed and engineered a complete system, specifically focusing on
high throughput plant imaging. This hardware-software product is a novel
{\em ``Portable Modular System for Automated Image Acquisition''} in the lab
and for certain field experiments. Our model system is described in
\cite{aa2009}. The algorithms below are part of ``The Image Analyzer''
component of our system. It is an object-oriented software application that
can be modified to accommodate essentially any reasonable analysis
scenario. The Image Analyzer package is developed in Linux, and it is ready
for experimental protocols by the biologist, see \cite{cyplant}.
This platform can provide high throughput results in a range of resolutions,
and accommodates highly flexible experimental protocols. This article
demonstrates the feasibility of using our system towards fully automated
high throughput phenotyping in functional genomics and systems
biology (Figure \ref{fig1}).

\subsection{Automated Preprocessing}

In earlier contributions by the senior author and Liya Wang (\cite{WangAS08},
\cite{lw2009}) image analysis and midline extraction were only
partially automated and required {\em ``preprocessing tasks''} prior to
application of the software. Preprocessing steps are highly data-dependent
image processing steps. In this article, we report successful automation of the
preprocessing, and a novel collection of algorithms that are also amenable
to generalization beyond the plant roots, as well as parallelization. The
preprocessing methods are adapted from earlier seminal research by Osher et
al (\cite{rl1992,aa2010}, called the Total Variation regularization (TV)
method. We remark that comparable approaches have been also treated in
literature using Bounded Variation (BV). As expected, direct application of
one set of programs to another typically shows varying degrees of success.
Therefore, it is anticipated that generalization of the automation
preprocessing would depend on the setting for image acquisition and would
vary in degrees of successful analysis in plant growth dynamics of various
kinds. The hardware and automation offered by CyPlant Solutions Inc. have a
standardized setting that removes a series of technical issues in order
to ensure the quality of data for the same genre of mathematical treatment
as in our algorithms and software. Our work in progress will provide a
critical examination of a number of algorithms and methods in literature,
and in appropriate cases, redesigns the needed image preprocessing algorithms
from scratch to be optimal for the tasks, such as those in CyPlant Solutions
standardized technologies.

We next present the outline of the algorithm. Let the image for one
gray-scale frame be represented by 
$$
	f: J \to \mathbb{R}.
$$
%
%
The basic idea is as follows: the typical image $f: J \to \mathbb{R}$
is written as an outcome of an unknown convolution kernel $K$ with compact 
support (e.g. a discrete Gaussian), applied to the original image
$\varphi$ (or a desirable model that in our case must be suitable for
segmentation and midline recovery of the branched structures,
cf. \cite{lw2009}), and an additive noise $\nu$, that we propose to model as
a Gaussian white noise, such that 
$$
	f=K * \varphi+\nu.
$$
The automation algorithm selects the kernel and a suitable model
for noise (that we must confront in practical applications), which will be
derived from the High Throughput Imaging System outputs in lieu of the
simplifying assumption above. We will begin with the constrained minimization
problem with appropriate norms, without further mention of simplicity.

For a class of images obtained by our method, the higher resolution
allowed us to select a delta function, and use the simpler form. However,
in the case of animal behavior, blurring occurs due to movement of the
animal or temporal loss of the automated focus. In the case of plants,
blurring occurs due to accumulation of moisture on inside surfaces of the
Petri dish covers. This means that we must use the general form below:
$$
	\varphi=\argmin_{\phi} \left\{ 
	F(\phi):= \frac{1}{2} \| K*\phi-f\|^2_{L^2}+ c_1\|\phi\|_{BV} \right\}.
$$
Here, $c_1$ is a scaling constant that is the trade-off between noise and
the desired image quality, such as having sharp edges in images of root
systems, and arg-min is over all images $\phi$, where $\varphi$  is the
desired form of the image suitable for algorithms of \cite{lw2009, so2005,
cv1998, tfc1998} (for an  iterative regularization to recover finer
scales). An intermediate problem that we solved numerically was the parallel
implementation of methods based on the following idea that is adapted from
a simplified mathematical result by Chan-Wong-Kaveh-Osher et al. that
amounts to a constrained optimization problem with $F$ as a function of
joint variables $(\phi,K)$:
\begin{align*}
(\varphi,\K)=\argmin_{\phi,K} & \bigg\{ F(\phi,K) := 
\frac{1}{2} \| K*\phi-f\|^2_{L^2} \\
&+ c_1\|\phi\|_{BV} + c_2 \|K\|_{BV}\bigg\}.
\end{align*}
When we view $F$ as a one-variable functional depending only on $K$ or only
$\phi$, by fixing the other one, then $F$ is convex,
while  $F(\phi,K)$ fails to be jointly convex. To overcome this
difficulty, we followed \cite{so2005} and instead numerically solved the two
Euler-Lagrange equations, where the $L^2$-conjugates are denoted by the
``hat'':
\begin{align*}
F_{\phi}(\phi,K) &=\hat K*(K*\phi-f)
	- c_1 \nabla \frac{\nabla \phi}{|\nabla \phi|}=0, \\
F_{K}(\phi,K) &=\hat \phi*(\phi*K-f)
	- c_2 \nabla \frac{\nabla \phi}{|\nabla \phi|}=0.
\end{align*}
Fixing $K$ first, we solve for $\phi$ from the first equation and then
switch the roles of $K$ and $\phi$ and exchange the two equations.
For a class of images according to our protocol, the higher resolution
allows one to select an iterative scheme that mimics convergence to the
appropriate delta function, which in turn allows one to use the simpler
form mentioned above. However, in the case of movement of the entire root
by sliding on the agar surface, blurring sometimes occurs; this could also
be a result of temporal loss of the automated focus. As mentioned before,
in the case of plants, blurring mostly occurs due to accumulation of
moisture on inside surfaces of the Petri dish covers. In our companion article
\cite{jt2010}, and in our 2008 WorldComp article \cite{WangAS08}, and
subsequent developments \cite{aa2010}, \cite{htd2008}, \cite{ivu2009}, we
have developed a set of new algorithms and their object-oriented C-code for
massively parallel-distributed hardware. This part of research was done in
collaboration with nVIDIA and SUN Network.com.

\subsection{Classification through Machine Learning}

Having quantified morphological features from the data set of movies, we
continue to discover candidates to serve as phenotypic traits that carry
the plant genotypic signature that account for distinguishing between the
wild type and the mutants. A number of these features depend on computation
of the midline of root images as argued in \cite{WangAS08}. The algorithm for
midline tracing in the present work is different from the main algorithm in
\cite{WangAS08}. In fact, the second generation of midline tracing
algorithms used in the present article is more robust, more accurate and much
faster \cite{jt2010}. We computed midlines from different regions of a root
image as follows.

\begin{figure}[htb]\centering
\includegraphics[width=\columnwidth]{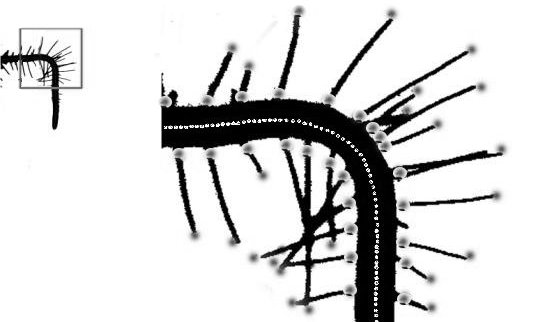}
\caption{\small Midpoints of a sample hook region. To avoid ambiguity of the
image in hair areas, we just indicate the start and the end points of the
hairs. For the hook region the midpoints are also shown.} \label{fig2}
\end{figure}

Gravitropism results in anisotropic expansion of cell walls and an uneven
change in the distribution of cells in the epidermis of the root in such a
way that the root growth tends to generate the observed bending with
respect to its growth in the original vertical direction \cite{baskin2005}.
The part of root before the bending region is referred to as the
{\em horizontal region}. The bending region itself is called the
{\em hook region} and finally, the part of the root after the hook region
is called the {\em vertical region}. Lengths
of these regions are calculated using the midline in these regions, and it
serves as a feature. Besides the length associated to the above-mentioned
regions, the curve made by the midline carries the information about the root
curvature during its gravitropic growth. Further considerations (omitted here)
indicate that the number of segments account for the changes in the growth
direction which could potentially be a significant representative feature.
The greatest change in the growth direction in the hook region is called the
{\em hook angle}. In addition to the geometric features of the midline of the
primary root growth, we also developed algorithms to extract growth
information regarding the root hairs. Discussion of the biological
significance and the algorithms for root hair growth information are
available in \cite{hd1,hd2}. Let us only mention that the number of root
hairs for each region of the root is also a feature that we added to
the list of morphological features for the classification purposes. Using
the root hair information, we formulate additional morphological features
which we call {\em dynamic features} because they correspond to the growth
velocity and growth acceleration of the primary root and the root hairs.
Figure \ref{fig2} shows these extracted features.

\noindent
All of the following features highlighted in the figure are considered as
representative features of the growth images: 
\begin{itemize}
	\item length of the vertical region,
	\item length of the horizontal region,
	\item length of the hook region,
	\item the number of segments,
	\item the hook angle,
	\item the average root hair length,
	\item the number of root hairs,
	\item the root growth velocity,
	\item root growth acceleration,
	\item hair growth velocity,
	\item hair growth acceleration, and
	\item hair density.
\end{itemize}

From a human observer's standpoint, the slow rate of growth of the root
morphological features results in ``invisible phenotypic traits'' that belong
to the complex dynamical system underlying the development and growth. In
\cite{hd2} we have developed the preliminary steps towards the in-depth study
of plant development within the framework of complex dynamical systems.
For comprehensive details of these algorithms readers are referred to visit our
website for \cite{jt2010,hd2,hd3,hd4,hd5} and \cite{hd6}. 

To recap, we have used image analysis algorithms and mathematical modeling
of dynamics of morphological features in order to isolate a number of
quantitative morphological and dynamical features, and using these
features, every movie from the growth process is assigned a
representation by the appropriate vector, as described earlier. The final
step towards discovery of quantitative phenotypic traits requires machine
learning for classifying the above-mentioned features, and to train the
machine through a set of so-called `training samples' or `examples' in
order to extract the appropriate weighted combination of features that
could carry genotypic signatures of the mutation (here, {\em mdr1}, which
is one of the members of the {\em mdr} gene family, is knocked to provide the
mutant). The machine learning method of choice here comes from the branch of
statistical learning theory, called support vector machine (SVM), see
\cite{vv2000,vv1998}. The flexibility of the SVM theory has proved
advantageous here, because we tailored to our needs an RBF-based SVM, where
RBF refers to the "Radial Basis Functions" as our choice for the SVM kernel
function. We applied the RBF-SVM method to 281 seedling growth movies (161
mutant {\em mdr1} and 120 wild type seedlings) where 12 features captured
them as described before. We used the MATLAB-SVM classifier \cite{matlab} and
calculated the precision of our classification outcome. Table \ref{tab1}
shows the results for each class of movies.
\begin{table}[htb]\centering
\begin{tabular}{lcc} 
                    & Magnitude of random        & \\
                    & selected training data set & Precision \\ \midrule
Class 0 (Mutant)    &         60                 & 98.8 \\ \midrule
Class 1 (Wild Type) &         45                 & 86.4 \\ \bottomrule
\end{tabular}
\caption{\small This table shows the result of applying the SVM method on
the data set, where the radial basis function was used as the kernel function.
Precision of the results shows efficiency of using the defined features
for capturing different classes of genotypic attitudes.} \label{tab1}
\end{table}

\noindent
The precision was computed via the standard error formula
$$
\frac{\text{true positive}}{\text{true positive} + \text{false positive}}.
$$ 

\begin{figure}[htb]\centering
\includegraphics[width=\columnwidth]{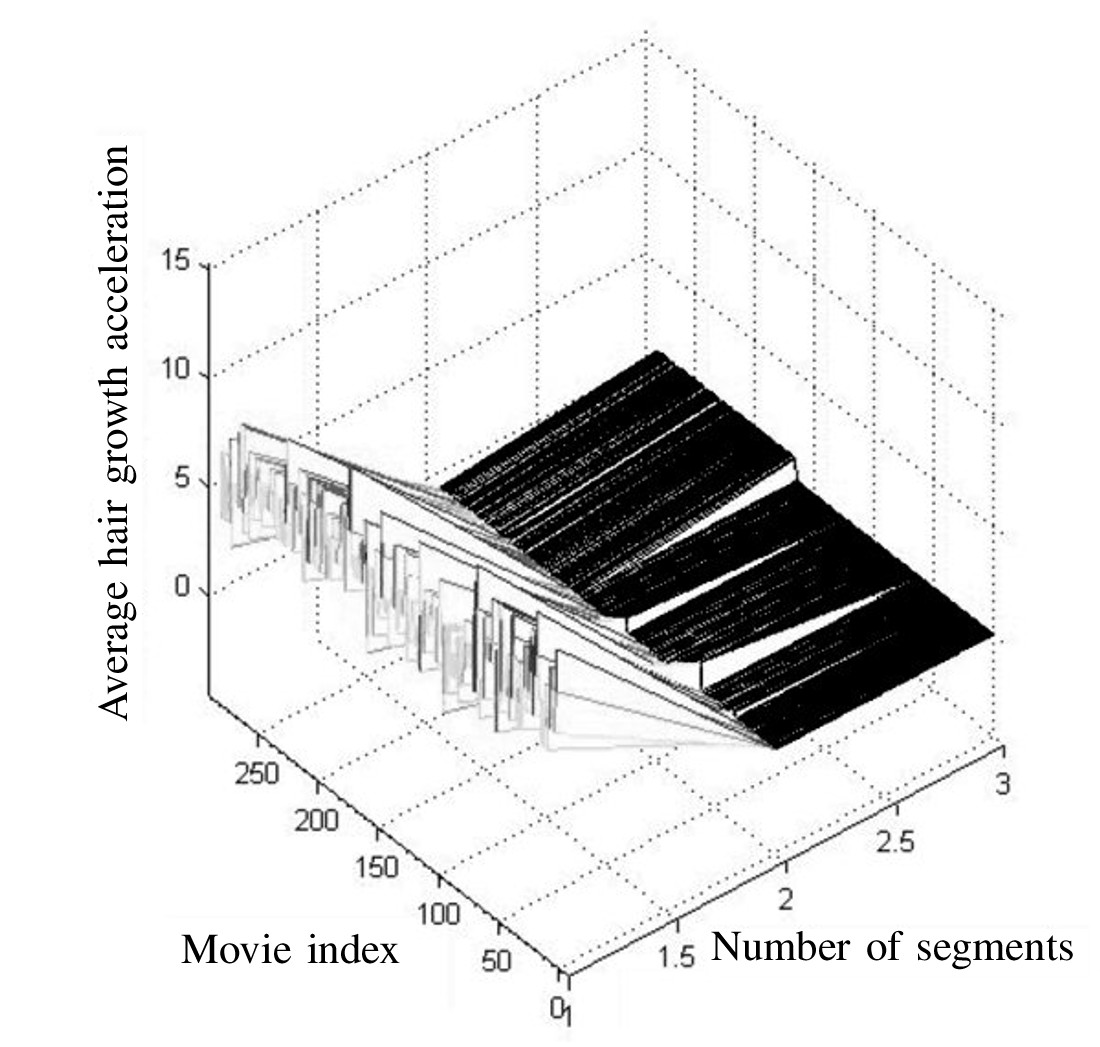}
\caption{\small
Projection of the examined movies on two features, ``number of segments'' and
``average hair growth acceleration''. Points are colored based on class ID;
black color is Class 0 and shades of gray are Class 1.} \label{fig3}
\end{figure}

Since twelve features were employed we could not show a full
$12$-dimensional figure of the classes. Instead, the projection of the movies
on the two coordinates corresponding to the features ``Number of segments''
and ``Average hair growth acceleration'' is shown in Figure \ref{fig3}.

\section{Discussion and Related Research}
Among higher plants, the model organism \Arabidopsis\ has been
studied in detail. As its genome is sequenced and better understood than
other similar plants, there are many opportunities for investigation of the
genotype-phenotype mapping. The development of the Arabidopsis root system
is a centerpiece in plant biology. Due to the wide availability of
a large amount of literature on root development, the Arabidopsis root system
continues to serve as an excellent model organ to investigate systems
biology of higher plants. The functional landscape of gene-protein network
dynamics is believed to be responsible for the regulation of growth and
development of roots. Multi-scale mathematical modeling of root growth
provides a critical element for a systematic study of the mechanisms
of regulatory transcription networks that operate on different scales,
specific interactions of numerous proteins and tightly intertwined protein
interaction networks. Gene expression studies need to be performed at
temporal and spatial resolutions (minutes and micrometers) that are
relevant to the dynamics of gene and phenotype crosstalk. Historically,
tropic growth responses have been at the center of such activities. Tropic
responses allow plants to redirect their growth in response to their
surrounding environment. The temporal dynamics of root growth is relatively
accurately measured by following the displacement of features at
extremities, such as the tips of roots and root hairs. It is necessary to
obtain velocity profiles of neighboring elements as they are moved by
expansion \cite{re1976,ws1992}. Velocity profiles can be obtained by imaging
a growing organ over time which provides measures of the position of the
externally applied marks, thus computing velocity as a function of position
\cite{hi1991}.

The mathematical analysis of image sequences goes back to numerous early
investigators, e.g. \cite{cf1979, bh1981}, and algorithmic methods substitute
for laborious and often subjective manual measurements
(\cite{al1995, jb1997, ds1998, aw2002}).
Several improvements were made by developing algorithms that measure the
spatial growth profiles using image-processing techniques that could also
utilize cell borders, intercellular air spaces or other physiological
structures that are followed throughout a sequence of images (see
\cite{bj1997, hn2000, ls2001, ak2006}).
Concepts originally developed by \cite{jm1998, jm2000} and
\cite{cw2002, cw2002b} and improved by \cite{ak2006} establish quantitative
relationships for curvature production and curvature angle distribution.
Nevertheless, there is a continuing need to develop better algorithms that
work more accurately in automated high throughput imaging systems such as
our system outlined above.

\section{Conclusion}
High throughput plant imaging systems require appropriate software
applications for accurate automated image analysis. We have developed a
prototype hardware-software product that operates as a {\em ``Portable Modular
System for Automated Image Acquisition and Analysis''} in the lab and some
field applications. This article provides evidence for feasibility of using
our system towards fully automated high throughput phenotyping in
challenging functional genomics and systems biology applications. The
methods described above also point to the many opportunities that machine
learning could offer plant functional biology, such as distinguishing
mutant from wild type seedling plants in gravitropism experiments. As we
observed, these applications could be adapted for other experimental
protocols in plant biology that attempt to quantify subtle phenotypic
traits in order to decipher the functions of genes-proteins. The general
problem of quantifying {\em all} phenotypic traits (for example, in tropism)
for the development and growth of the plant root system remains a formidable
challenge that is certain to inspire new insights in machine learning and
analysis of massive biological imaging data.





\end{document}